# Anti-lock Brake System for Integrated Electric Parking Brake Actuator Based on Sliding-mode Control

## Abstract

Integrated electric parking brake (iEPB) is popularizing on passenger cars due to its easier operation and automatic functions. As a parking brake, EPB have to act as the secondary brake system in case of hydraulic brake failure. To guarantee the stability and safety of a car during iEPB braking, the rear slip ratio has to be controlled accurately within the optimized value to get the shortest brake distance without undesired loss of control. In this paper, a sliding-mode controller (SMC) is investigated to achieve rear-wheel anti-lock brake control, which is robust against uncertainties and disturbance of the parameters. And a sliding-mode observer (SMO) is present to estimate the load torque of d.c. motor and calculate the brake torque. The tyre/road friction coefficient estimator is designed to obtain the optimal rear slip ratio timely. The simulation model of iEPB system is initially constructed in AMESim and the vehicle model is built in MATLAB/Simulink, and the complete system is co-simulated by these two software simultaneously with different road conditions. Simulation results show that the proposed observer and estimator are feasible. This study may provide a useful method to realize rear slip ratio control so that the safety and stability of vehicle could be improved significantly in specified condition

## 1. Introduction

iEPB has superior performance of easier operation and automatic functions comparing with the conventional lever parking brake system.[1,2] so it has been widely used in modern vehicle and the conventional will be replaced fully in the near future. Another important advantage of iEPB is that the information is shareable with other brake-by-wire system by controller area network (CAN) [3].

Not only can iEPB be regarded as a parking brake, but also it has to act as the secondary brake when the conventional hydraulic brake system fails. However, iEPB is flawed in process of as secondary brake. Rear slip ratio control hasn't been realized. Therefore, vehicle may generate the risk of spin when the secondary brake is activated without slip ratio control. [4,5] For this problem, it hasn't been reported in earlier references. To guarantee the stability and safety of a car during iEPB braking, the rear slip ratio should be controlled accurately and smoothly to reach optimal slip ratio and get shortest brake distance on the basis of vehicle dynamic control [6, 7, 8].

As we known, the slip ratio control is that adjusting brake torque to change velocity of wheels and prevent the tyre locking essentially. At present, many methods to control slip ratio have been discussed. Fan X et al. [9] and El-Garhy A M et al. [10] suggested fuzzy logic algorithm for vehicle stability control system. Li J et al. [11] discussed three anti-lock braking control algorithms which are PID, logic threshold and sliding mode variable structure [12]. Mirzaeinejad H et al. [13] researched the scheduled optimal control algorithm for ABS system and Savaresi S M et al. [14] applied mixed slip-deceleration control in automotive braking systems. In addition, Pasillas-Lépine W et al. [15] presented the nonlinear wheel slip control algorithm with simulation and experimental validation. For the above researches, the optimal slip ratio should be given and the brake torque should be known beforehand. However, the brake torque cannot be measured directly due to cost limit and small installation space. But, it can be calculated by the observed load torque of d.c. motor. Moreover, the optimal slip ratio varies with the tyre/road friction coefficient, which relates to the road conditions directly. The road conditions are usually estimated with different estimators and the optimal slip ratio can be obtained.

For the tyre/road friction coefficient estimation, Rajamani R et al. [16] and Chen Y et al. [17] gave the recursive least-squares parameter identification method to estimate the individual wheel tyre/road friction coefficient. Faraji M et al. [18] designed an optimal pole-matching observer for estimating the tyre/road friction force. Wang R et al. [19] and Ko S et al. [20] proposed the friction coefficient estimation through the integrated longitudinal force and lateral force, respectively. Rath J J et al. [21] studied the combination of nonlinear Lipschitz observer to estimating road frictional condition and Castillo J J et al. [22] presented an extended Kalman filter (EKF) tyre friction coefficient estimation method. In addition, Veluvolu K C et al. [23] proposed an SMO for the tyre friction estimation. Based on these studies, the tyre/road friction coefficient can be estimated and the optimal slip ratio can be obtained to as the control target.

As for how to get load torque of d.c. motor, many observation methods are investigated as follows. Leu V Q et al. [24] presented a load torque observer based on the linear matrix inequality (LMI). Chen J F et al. [25] exploited a state observer is described based on square-root unscented Kalman filtering. Hong M et al. [26] studied the nonlinear torque observer and Li L et al. [27] designed a sliding-mode observer to estimate the load torque. For the above discussion, the brake torque calculated with the estimated load torque of d.c. motor is necessary, and there will be not effective without the load torque of d.c. motor information.

According to analysis for studies about vehicle ABS control method, a SMC algorithm is presented in this paper. the rear slip ratio and brake torque can be controlled based on the load torque of d.c. motor observed by SMO and the tyre/road friction coefficient estimated using the *μ-slip* curve for iEPB system. Finally, it is worth mentioning that the iEPB actuator is constructed in AEM-Sim, the difficulty of building a complex system model is reduced greatly and the model is closer to real.

The paper is organized as follows. In the second section, the vehicle model and the tyre model are built in MATLAB/Simulink, and the dynamic model of iEPB actuator is constructed in AEM-Sim. In the third section, the derivation process of observers is introduced in detail. In the fourth section, SMC is designed and analyzed. In the fifth section, the complete system model is co-simulated between AME-Sim and MATLAB/Simulink, the simulation results are illustrated. The conclusions are summarized in the sixth section.

## 2. Structures and models

### 2.1 Vehicle model

The vehicle is regarded as a rigid body, only the longitudinal and pitch dynamics are considered for the rear-wheel anti-lock brake control system. Without loss of generality, the air drag resistance, the rolling resistance and viscous resistance of wheel are ignored because of space constraints. A six-degree-of-freedom dynamics model is built, including the rotation of each wheel ($\omega_{fl}$, $\omega_{fr}$, $\omega_{rl}$, $\omega_{rr}$), longitudinal motion ($v_x$) and pitch motion, as shown in Figure 1. The vehicle dynamic equations are expressed as follows.

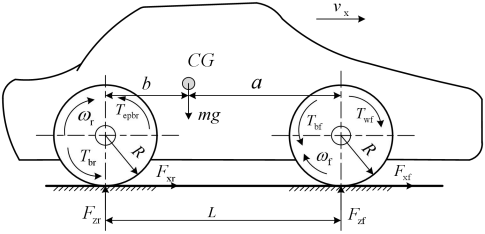

Figure 1. The six-degree-of-freedom vehicle model

The equation of the longitudinal motion is

$$m\dot{v}_x = \begin{pmatrix} F_{xfl}(\lambda_{fl},F_{zfl}) + F_{xfr}(\lambda_{fr},F_{zfr}) \\ + F_{xrl}(\lambda_{rl},F_{zrl}) + F_{xrr}(\lambda_{rr},F_{zrr}) \end{pmatrix}$$
$$= \sum F_{xij}(\lambda_{ij},F_{zij})$$

(1)

where the subscript $f$ and $r$ represent front wheel and rear wheel respectively and the subscript $l$ and $r$ represent left wheel and right wheel separately. The subscript $i$ is front or rear wheel and the subscript $j$ is left or right wheel. $m$ is the vehicle mass, $F_{xij}(\lambda_{ij},F_{zij})$ is the longitudinal tyre force which is described in Equation (4), $F_{zij}$ is the vertical load which is expressed in Equations (2) and $\lambda_{ij}$ is the slip ratio.

In the process of deceleration or acceleration, the load transfer of vehicle and the effect of the body pith motion are considered. So the vertical loads on front and rear axle is given as

$$\begin{cases} F_{zf} = \dfrac{m}{2L}(gb - \dot{v}_x h_g) \\ F_{zr} = \dfrac{m}{2L}(ga + \dot{v}_x h_g) \end{cases}$$

(2)

where $b$ is the distance between the gravity center and rear axle, $a$ is the distance between the gravity center and front axle, $h_g$ is the height of the gravity center and $L$ is the distance between the front and rear axles.

The equations of wheel rotational motions are

$$\begin{cases} J_f \dot{\omega}_{fl} = F_{xfl}(\lambda_{fl},F_{zfl})R_f - T_{bfl} + T_{wfl} \\ J_f \dot{\omega}_{fr} = F_{xfr}(\lambda_{fr},F_{zfr})R_f - T_{bfr} + T_{wfr} \\ J_r \dot{\omega}_{rl} = F_{xrl}(\lambda_{rl},F_{zrl})R_r - T_{brl} - T_{epbrl} \\ J_r \dot{\omega}_{rr} = F_{xrr}(\lambda_{rr},F_{zrr})R_r - T_{brr} - T_{epbrr} \end{cases}$$

(3)

where $J$ is rotational inertias of the front and rear wheels, $R$ is the radiuses of the front and rear wheels, $T_b$ are the hydraulic braking torques of wheel, and $T_{epb}$ is the braking torques of rear wheel from iEPB system, and $T_w$ is driving torques of front wheel.

## 2.2 Tyre model

The magic formula (MF) is used to describe nonlinear characteristics of tyre forces and is also accurate in off-line simulation. Assume that the vehicle drive straightly, tyre longitudinal force $F_x$ can be described as:

$$F_{xij} = D\sin\left\{C\arctan\left(B\lambda_{ij} - E(B\lambda_{ij} - \arctan B\lambda_{ij})\right)\right\}$$

(4)

where $B$ is the stiffness factor of the longitudinal force coefficient curve, $C$ is the shape factor of the longitudinal force coefficient curve, $D$ is the peak value and $E$ is the curvature factor of the longitudinal force coefficient curve which describes the curve shape near the maximum longitudinal force coefficient.

Factors of longitudinal force are

$$\begin{cases} C = a_2 \\ D = a_0 F_z^2 + a_1 F_z \\ B = \dfrac{a_3 F_z^2 + a_4 F_z}{CDe^{a_5 F_z}} \\ E = a_6 F_z^2 + a_7 F_z + a_8 \end{cases}$$

(5)

where $a_0$-$a_8$ are parameters from bench tests. Their values of the Michelin tire Michelin MXV8 are shown in Table 1. When the wheel load is a fixed value, the actual relationship between the tyre longitudinal force and the slip ratio under different road conditions is similar to that shown in Figure 2.

| $a_0$ | $a_1$ | $a_2$ | $a_3$ | $a_4$ | $a_5$ | $a_6$ | $a_7$ | $a_8$ |
|---|---|---|---|---|---|---|---|---|
| 0 | 1000 | 1.55 | 60 | 300 | 0.17 | 0 | 0 | 0.2 |

Table 1. Tyre's parameters (Michelin® MXV8 205/55R16 91V).

The Magic Formula is very complex, so the look up table method with the experimental data based on the Magic Formula is adopted to reduce the computing burden. The look-up table method can be given as

$$F_{xij} = G\left(F_{zij}, \mu_{ij}, \lambda_{ij}\right)$$

(6)

where $G$ is the function of longitudinal force $F_{xij}$ with vertical load $F_{zij}$, road friction $\mu_{ij}$ and slip ratio $\lambda_{ij}$.

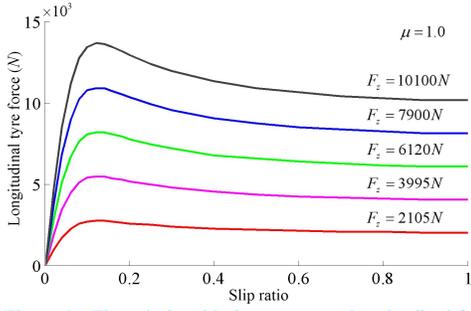

Figure 2. The relationship between tyre longitudinal force and slip ratio

When the vehicle is running, the slip ratio is defined as:

$$\lambda_{ij} = \begin{cases} \dfrac{v_x - \omega_{ij} R_i}{v_x} & (v_x \geq \omega_{ij} R_i, v_x \neq 0) \\ \dfrac{\omega_{ij} R_i - v_x}{\omega_{ij} R_i} & (\omega_{ij} R_i \geq v_x, \omega_{ij} R_i \neq 0) \end{cases}$$

(7)

## 2.3 Brake system model

### Structure and characteristics of the iEPB system

The iEPB is a electromechanical system. Its actuator is composed of a d.c. motor, a synchronous belt, a reduction gear, a screw-nut pair, a caliper and a piston, as shown in Figure 3. The working principle can be described as follows: an electric control unit (ECU) is responsible for the signals and directives; an H-bridge composed by 4 MOSFETs is used to drive the motor; pulse width modulation (PWM) control can adjust the d.c. motor voltage to make iEPB actuator apply and release brake force.

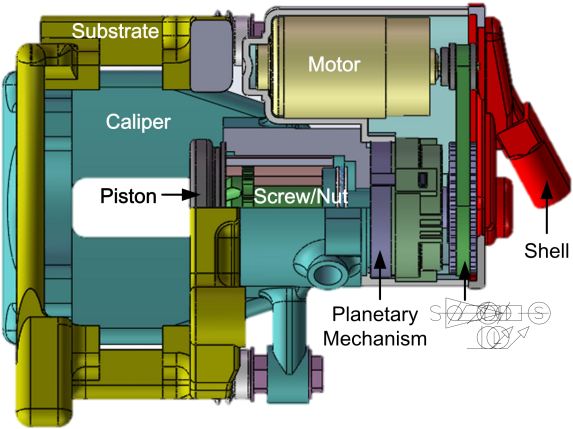

Figure 3. The structure of the novel EPB actuator

### Mathematical model of iEPB

#### D.C. motor model

The d.c. motor of the iEPB can provide the original torque to drive the system by converting electrical energy into mechanical energy. Its parameters are listed in Table 2.

Based on the Kirchhoff voltage law, the electrical equation of the d.c. motor is

$$\begin{cases} uV_a = E_b + R_a i_a + L_a \dot{i}_a \\ E_b = k_e \omega_m \end{cases}$$

(8)

where $u$ is the duty of PWM, $V_a$ is the battery voltage, $i_a$ is the armature current, $L_a$ is the armature inductance, $E_b$ is the back electromotive voltage, $R_a$ is the armature resistance, $k_e$ is the motor electric constant and $\omega_m$ is the speed of the motor.

The system motion status is described by mechanical equation.

$$\begin{cases} J_m \dot{\omega}_m = T_a - c_m \omega_m - T_m \\ T_a = k_t i_a \end{cases}$$

(9)

where $J_m$ is the rotational inertia of d.c. motor, $T_a$ is the generated torque from the motor, $T_m$ is the external load torque in the motor axis, $c_m$ is the damping coefficient of the motor and $k_t$ is the torque constant, $k_t$ is equal to $k_e$ in consistent units.

Table 2. Parameters of the d.c. motor

| Symbol | Descriptions | Unit | Value |
|---|---|---|---|
| $U$ | Rated voltage | V | 12 |
| $I_a$ | Rated current | A | 4.34 |
| $n$ | Rated revolution speed | rpm | 8944 |
| $P$ | maximum power | W | 85 |
| $L$ | Stator inductance | H | 0.00083 |
| $R$ | Stator resistance | Ω | 0.365 |
| $k_e$ | Back emf coefficient | V/rpm | 0.11 |
| $k_t$ | Torque coefficient | N m/A | 0.11 |
| $J_m$ | Moment of inertia of rotor | Kg m$^2$ | 4.21×10$^{-3}$ |
| $B_m$ | Viscous damping coefficient | N m s/rad | 0.001 |

*Screw-nut mechanism model*

In iEPB system, the screw-nut mechanism is used to convert the rotational motion of the d.c. motor to the linear motion of the piston. It can be described in Equation (10). The nut moves forward and backward to apply and release brake force. Furthermore, one important characteristic of the mechanism is that the motor power could be cut off when the needed braking force is reached. Then the clamping force could be applied steadily due to the screw-nut self-locking characteristic.

$$\theta_{nut} = \frac{2\pi s_{nut}}{P_h} \tag{10}$$

where $\theta_{nut}$ is the rotational angle of screw, $s_{nut}$ is the displacement of nut and $P_h$ is the pitch of screw-nut.

In order to build the mathematic model of the screw-nut mechanism precisely, the relative motion is regarded as a slide block above an inclined plane. The force diagram of the screw-nut pair is analyzed, as shown in Figure 4.

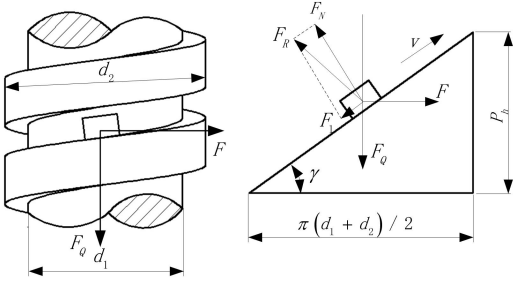

Figure 4. Force diagram between screw and nut

where $d_1$ and $d_2$ are the inner diameter and the outer diameter of the screw, $\gamma$ is the thread angle, $F$ is the horizontal drive force, $F_Q$ is the load force in screw's axis direction, $F_{fc}$ is the friction force between the screw and the nut, $F_N$ is the normal force, and $v$ is the linear velocity of the nut along with the inclined plane.

So the force equations are set up as follows.

$$\begin{cases} T_{screw} = Fd/2 \\ F_{fm} = F\cos\gamma - F_Q\sin\gamma \\ F_{fc} = u_s\left(F\sin\gamma + F_Q\cos\gamma\right) \end{cases} \tag{11}$$

where $d$ is the medium diameter of the screw, $T_{screw}$ is the input torque, $F_{fm}$ is the friction force that is less than the maximum static friction force $F_f$ when the screw/nut is static, $u_s$ is the sliding friction coefficient.

Therefore the friction model of the screw/nut pair is described as follows.

$$F_f(v, F_{fm}) = \begin{cases} F(v) & \text{if } v \neq 0 \\ F_{fm} & \text{if } v = 0 \; \& \; |F_{fm}| < F_f \\ F_s \, \text{sgn} \, F_{fm} & \text{if } v = 0 \; \& \; |F_{fm}| \geq F_f \end{cases} \tag{12}$$

Where $F(v)$ is the sliding friction force when the screw-nut mechanism motions.

The friction force includes the Coulomb friction, the static friction, the viscous friction and the Stribeck friction. [28] According to the Olsson friction theory, the frictional function is written as

$$F(v) = F_{fc} + (F_s - F_{fc})e^{-\left|\frac{v}{v_s}\right|\delta_s} + \mu_v v \tag{13}$$

where $v_s$ is the Stribeck speed, $\delta_s$ is the experimental value and $\mu_v$ is the coefficient of viscous friction.

The d.c. motor's speed becomes smaller through the retarding mechanism, so the magnitudes of the screw-nut viscous friction and the Stribeck friction is much small compared with the Coulomb friction, these could be neglected in the mathematic model. Equation (13) could be reduced to

$$F(v) = F_{fc} = u_s\left(F\sin\gamma + F_Q\cos\gamma\right) \tag{14}$$

*Caliper mechanism model*

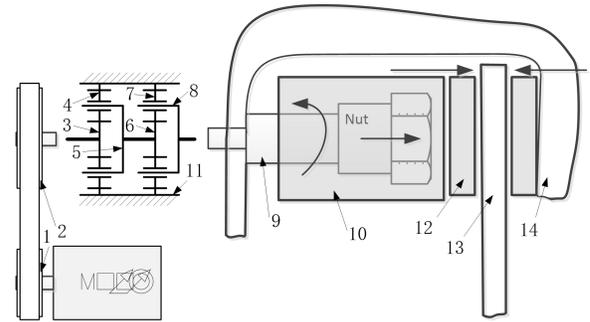

Figure 5. Schematic diagram of iEPB actuator

1.belt gear I 2.belt gear II 3.Solar gear I 4.Planetary gear I 5.Planetary carrier I 6.Solar gear II 7.Planetary gear II 8.Planetary carrier II 9. Screw 10.Nut 11.Gear rings 12.friction pads 13.Brake disc 14.Callipers

When brake force need to be applied, the clearance between friction plate and brake disc is compensated first, as shown in Figure 5. Then, the brake force is generated. The clamping force equation can be built as

$$F_Q = k_c\left(s_{nut} - s_{mc}\right) + b_c \dot{s}_{nut} \tag{15}$$

where $k_c$ is the stiffness coefficient of the caliper, $b_c$ is the damping coefficient of the caliper and $s_{mc}$ is the brake clearance between friction plate and brake disc.

### The complete transmission system model of iEPB

When the motor rotates forward, the kinetic equations of actuator are:

$$\begin{cases} T_m = T_1 = J_1\dot{\omega}_1 + c_1\omega_1 + \dfrac{T_2}{\eta_{12}i_{12}} \\ T_2 = T_3 = J_2\dot{\omega}_2 + J_3\dot{\omega}_3 + c_2\omega_2 + m_4r_5^2\dot{\omega}_5 + NJ_4\dot{\omega}_4 + \dfrac{T_5}{\eta_{35}i_{35}} \\ T_5 = T_6 = J_5\dot{\omega}_5 + J_6\dot{\omega}_6 + m_7r_8^2\dot{\omega}_8 + NI_7\dot{\omega}_7 + \dfrac{T_8}{\eta_{68}i_{68}} \\ T_8 = J_8\dot{\omega}_8 + c_3\omega_8 + J_9\dot{\omega}_9 + F_{fc}\dfrac{d}{2}\dfrac{1}{\eta_{89}} + \dfrac{F_Q P_h}{2\pi\eta_{89}} \end{cases}$$

(16)

When the motor rotates backward, the kinetic equations of actuator are:

$$\begin{cases} T_m = T_1, T_1 - J_1\dot{\omega}_1 - c_1\omega_1 = \dfrac{\eta_{21}T_2}{i_{12}}, T_2 = T_3 \\ T_3 + J_2\dot{\omega}_2 + J_3\dot{\omega}_3 + c_2\omega_2 + m_4r_5^2\dot{\omega}_5 + NJ_4\dot{\omega}_4 = \dfrac{\eta_{53}T_5}{i_{35}} \\ T_5 = T_6, T_6 + J_5\dot{\omega}_5 + J_6\dot{\omega}_6 + m_7r_8^2\dot{\omega}_8 + NJ_7\dot{\omega}_7 = \dfrac{\eta_{86}T_8}{i_{68}} \\ T_8 + J_8\dot{\omega}_8 + c_3\omega_8 + J_9\dot{\omega}_9 + F_{fc}\dfrac{d}{2}\dfrac{1}{\eta_{98}} = \dfrac{F_Q P_h}{2\pi\eta_{98}} \end{cases}$$

(17)

where $T_k$ is the torque of part $k$, $J_k$ is the inertia of part $k$, $\omega_k$ is the rotating speed of part $k$, $m_k$ is the mass of part $k$, $r_k$ is radius of part $k$, $\eta_{nk}$ is the transmission efficiency from part $n$ to part $k$, $N$ is the quantity of planetary gears, $i_{nk}$ is the transmission ratio from part $n$ to part $k$, $c_1, c_2$ and $c_3$ are the rotating damping coefficients of input shaft bearing, planetary gear's shaft bearing, and output shaft bearing.

$$J_{eq} = \begin{cases} J_1 + \dfrac{1}{\eta_{12}}[\dfrac{J_2+J_3}{i_{12}^2} + \dfrac{Nm_4r_5^2}{i_{12}i_{15}} + \dfrac{NJ_4}{i_{12}i_{14}} + \dfrac{1}{\eta_{35}}(\dfrac{J_5+J_6}{i_{12}i_{35}i_{16}} + \\ \qquad \dfrac{NJ_7}{i_{12}i_{35}i_{17}} + \dfrac{Nm_7r_8^2}{i_{12}i_{35}i_{18}} + \dfrac{1}{\eta_{68}}\dfrac{J_8+J_9}{i_{12}i_{35}i_{68}i_{18}})], (\omega_m > 0) \\ J_1 - \eta_{21}[\dfrac{J_2+J_3}{i_{12}^2} + \dfrac{Nm_4r_5^2}{i_{12}i_{15}} + \dfrac{NJ_4}{i_{12}i_{14}} + \eta_{53}(\dfrac{J_5+J_6}{i_{12}i_{35}i_{16}} + \\ \qquad \dfrac{NJ_7}{i_{12}i_{35}i_{17}} + \dfrac{Nm_7r_8^2}{i_{12}i_{35}i_{18}} + \eta_{86}\dfrac{J_8+J_9}{i_{12}i_{35}i_{68}i_{18}})], (\omega_m < 0) \end{cases}$$

(20)

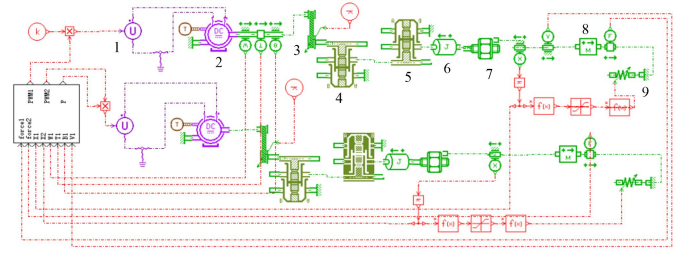

Figure 6. The co-simulation model between AME-Sim and Simulink

The friction force of gears is relevant to the transmitted torque, so the engaging efficiency is defined to include the influence of friction force.

$$\begin{cases} \dfrac{\omega_m}{i_{12}} = \dfrac{\omega_1}{i_{12}} = \omega_2 = \omega_3, \omega_4 = (1+\dfrac{1}{i_{34}})\omega_5 - \dfrac{1}{i_{34}}\omega_3 = \dfrac{\omega_1}{i_{14}}, \\ \omega_6 = \omega_5 = \dfrac{\omega_3}{i_{35}} = \dfrac{\omega_1}{i_{16}}, \omega_7 = (1+\dfrac{1}{i_{68}})\omega_8 - \dfrac{1}{i_{68}}\omega_6 = \dfrac{\omega_1}{i_{17}}, \\ \omega_8 = \dfrac{\omega_6}{i_{68}} = \dfrac{\omega_1}{i_{18}}, i_{12} = \dfrac{Z_2}{Z_1}, i_{35} = 1+\dfrac{Z_{11}}{Z_3} = \dfrac{1}{i_{53}}, i_{68} = 1+\dfrac{Z_{11}}{Z_6} = \dfrac{1}{i_{86}} \end{cases}$$

(18)

According to Equations (16)-(18), the equivalent moment of inertia $J_{eq}$ and the equivalent rotating damping coefficient $c_{eq}$ are obtained as

$$c_{eq} = \begin{cases} c_1 + \dfrac{c_2}{\eta_{12}i_{12}^2} + \dfrac{c_3}{\eta_{12}\eta_{35}\eta_{68}i_{12}i_{35}i_{68}i_{18}}, (\omega_m > 0) \\ c_1 - \dfrac{\eta_{21}c_2}{i_{12}^2} - \dfrac{\eta_{21}\eta_{53}\eta_{86}c_3}{i_{12}i_{35}i_{68}i_{18}}, (\omega_m < 0) \end{cases}$$

(19)

Then, Equations (16) and (17) can be simplified as:

$$T_m = T_r + J_{eq}\dot{\omega}_m + c_{eq}\omega_m$$

(21)

$$T_r = \begin{cases} \dfrac{\pi dF_f + F_Q P_h}{2\pi i_{12}i_{35}i_{68}\eta_{12}\eta_{35}\eta_{68}\eta_{89}}, (\omega_m \geq 0) \\ \dfrac{\eta_{21}\eta_{53}\eta_{86}(-\pi dF_f + F_Q P_h)}{2\pi i_{12}i_{35}i_{68}\eta_{89}}, (\omega_m < 0) \end{cases}$$

(22)

where $T_r$ is the resistance torque on the output shaft of d.c. motor.

So the equivalent state equations are

$$\begin{cases} \dot{\theta}_m = \omega_m \\ \dot{\omega}_m = -\dfrac{1}{J_n}T_r(\theta) - \dfrac{c_n}{J_n}\omega_m + \dfrac{k_t}{J_n}i_a \\ \dot{i}_a = -\dfrac{k_e}{L_a}\omega_m - \dfrac{R_a}{L_a}i_a + \dfrac{V_a}{L_a}u \end{cases}$$

(23)

where

$$\begin{cases} c_n = c_m + c_{eq} \\ J_n = J_m + J_{eq} \end{cases}$$

(24)

*Brake model*

When the hydraulic brake system fails in the process of diving, iEPB system is operated as the secondary brake. Hence, the brake torque is applied by only iEPB actuator.

In order to get more realistic brake model, it is proposed that the iEPB model is built in AEM-Sim on the basis of the structure of the iEPB system, as shown Figure 6. Because most of modules can be called directly without mathematical formulas, it's simpler and more convenient.

The relationship between the clamping pressure and brake torque can be defined as follows:

$$T_{epbrj} = \dfrac{\pi}{4}D^2 \mu_b r p_{epb}$$

(25)

where $D$ is the diameter of piston, $\mu_b$ is the friction coefficient between friction plate and brake disc, $p_{epb}$ is the brake pressure provided by iEPB actuator, and $r$ is the equivalent distance between the center of brake disc and friction plate.

The number designations in Figure 6 means that: NO.1 is the battery, NO.2 is the DC motor, NO.3 is the synchronous belt transmission mechanism, NO.4 is the first planetary gears and NO.5 is the second one, NO.6 is the inertia of the system, NO.7 is the screw-nut mechanism, NO.8 is the mass of the braking piston, NO.9 is the friction plate.

## 3. Important parameters estimation

### 3.1 Sliding-mode observer for load torque

Actually, the pressure sensor can't be installed near the brake disc because of limited space. Thus, the clamping pressure can't be measured. In addition, the clamping force is difficult to be calculated accurately. So a SMO is investigated to estimate the load torque of d.c. motor.

The design process of SMO is described in detail in this part. Firstly, considering that the sample time is very short, which is set 1ms in this paper, it is reasonable to assume that the load torque is constant in a control period, [29] the dynamic equation of d.c. motor could be rewritten as

$$\begin{cases} \dot{\omega}_m = \dfrac{k_t}{J_n}i_a - \dfrac{c_n}{J_n}\omega_m - \dfrac{1}{J_n}T_r \\ \dot{T}_r = 0 \end{cases}$$

(26)

Then, the motor rotational speed and load torque are considered as observed objects, the SMO could be established in Equation (27).

$$\begin{cases} \dot{\hat{\omega}}_m = \dfrac{k_t}{J_n}i_a - \dfrac{c_n}{J_n}\hat{\omega}_m - \dfrac{1}{J_n}\hat{T}_r + U \\ \dot{\hat{T}}_r = gU \end{cases}$$

(27)

$$U = k\,\mathrm{sgn}(\hat{\omega}_m - \omega_m)$$

(28)

where $k$ is the sliding mode gain, g is the feedback gain, $\hat{\omega}_m$ and $\hat{T}_r$ are estimated values of motor rotational speed and load torque respectively.

The error equations are

$$\begin{cases} \dot{e}_1 = -\dfrac{c_n}{J_n}e_1 - \dfrac{1}{J_n}e_2 + U \\ \dot{e}_2 = gU \end{cases}$$

(29)

$$\begin{cases} e_1 = \hat{\omega}_m - \omega_m \\ e_2 = \hat{T}_r - T_r \end{cases}$$

(30)

The error of rotational speed is chosen as sliding-mode surface, and then $s$ is

$$s = e_1 = \hat{\omega}_m - \omega_m = 0$$

(31)

According to sliding-mode control theory, the stability condition of observer is

$$\begin{aligned} \dot{V} &= s\dot{s} = e_1\dot{e}_1 \\ &= e_1\left(-\dfrac{c_n}{J_n}e_1 - \dfrac{1}{J_n}e_2 + k\,\mathrm{sgn}\,e_1\right) \\ &= -e_1\,\mathrm{sgn}\,e_1\left(\dfrac{c_n}{J_n}\dfrac{e_1}{\mathrm{sgn}\,e_1} + \dfrac{1}{J_n}\dfrac{e_2}{\mathrm{sgn}\,e_1} - k\right) \le 0 \end{aligned}$$

(32)

Obviously, $-e_1 \operatorname{sgn} e_1$ is no more than zero, so the item in the bracket need to be no less than zero. The range of the sliding-mode gain $k$ could be obtained.

$$k \leq \frac{c_n}{J_n} \frac{e_1}{\operatorname{sgn} e_1} + \frac{1}{J_n} \frac{e_2}{\operatorname{sgn} e_1}$$

(33)

$$k \leq (c_n/J_n)|e_1| - (1/J_n)|e_2|$$

(34)

When the observer comes into sliding mode, $s$ is zero which means $e_1 = \dot{e}_1 = 0$. Equation (29) could be simplified as

$$\dot{e}_2 - \frac{g}{J_n} e_2 = 0$$

(35)

Then, the estimation error of load torque is

$$e_2 = C e^{-\frac{g}{J_m} t}$$

(36)

where $C$ is a constant. To ensuring the stability of the system, $-g/J_m > 0$ should be satisfied which means $g < 0$. The decay rate of $e_2$ is determined by the value of $g$.

Generally, it is inescapable that there is an chattering phenomenon in sliding mode control, which has a direct effect on the system performance.[30] In order to eliminate the chattering, the sign function $sgn(s)$ is replaced by a piecewise linear approximation saturation function $sat(s)$.

$U$ can be described as

$$U = k\,sat(\hat{\omega} - \omega)$$

(37)

The saturation function is defined as

$$sat(\frac{s_r}{\phi}) = \begin{cases} -1, & s_r < -\phi \\ \operatorname{sgn}\left(\frac{s_r}{\phi}\right), & -\phi \leq s_r \leq \phi \\ 1, & s_r > \phi \end{cases}$$

(38)

where $\phi$ is the width of the boundary.

So the brake torque can be calculated as

$$T_{epbr} = \frac{2\pi \eta_{89} \mu_b r}{P_h} \hat{T}_r$$

(39)

### 3.2 Tyre/road friction coefficient estimator

To achieve the anti-lock brake technique for iEPB, the rear slip ratio is considered as the control goal. However, the desired slip is time varying because of road friction coefficient. So it is described that the road friction coefficient and the desired slip ratio are estimated at the control period $T$.

**Tyre/road friction coefficient estimation**

The tyre/road friction coefficient plays an important role in the process of braking. [31] The tyre force can be determined by the road friction, the control target of upper controller can be affected by it. Thus, the tyre/road friction coefficient should be identified accurately and rapidly. An estimation method is proposed that tyre/road friction coefficient is estimated by $u$-slip curve. The longitudinal tyre forces can be calculated from Equation (4) and the vertical loads of rear wheel can be got from Equation (2). So the utilized friction coefficient is

$$\mu_{xi}^t = \frac{F_{xi}^t}{F_{zj}^t}$$

(40)

Then, the change rate of $\mu_x$ in one control period $T$ is

$$k^t = \begin{cases} \frac{\mu_{xi}^t - \mu_{xi}^{t-1}}{\lambda^t - \lambda^{t-1}} & (\lambda^t \neq \lambda^{t-1}) \\ k^{t-1} & (\lambda^t = \lambda^{t-1}) \end{cases}$$

(41)

where $\mu_x^t$ and $\lambda^t$ are the utilized friction coefficient and slip ratio at one control period, respectively.

If $k^t \in [k_1 - \delta_1, k_1 + \delta_2]$, the tyre is working in the linear area, which means the tire force and the utilized friction coefficient increase with the brake pressure. The relationship between $\mu_x^t$ and $\lambda^t$ will change in the direction of $A$ to $B$, as shown in Figure 7. It's reasonable to define the current road friction coefficient as follows:

$$\mu_{x\max}^t = \mu_x^t + k_1(\lambda^t - \lambda^{t-1})$$

(42)

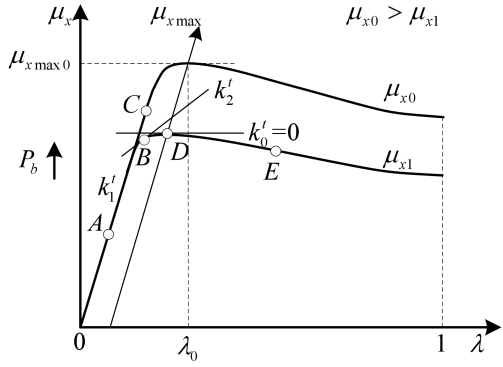

Figure. 7. Road friction estimation using $\mu$-slip curve

If $k^t \in [k_0+\delta_3, k_1-\delta_1)$, the tyre is working in the transitional area, which is stable. In this case, if the brake pressure increases, the tyre force and the utilized friction coefficient will increase too. The relationship between $\mu_x^t$ and $\lambda^t$ will change along the curve B to D, where the utilized friction coefficient is approximating the maximum value, as shown in Figure 7. The estimated road peak friction can be described as:

$$\mu_{x\max}^t = \mu_x^t + k^t(\lambda^t - \lambda^{t-1})$$

(43)

If $k^t \in [-\delta_4, k_0+\delta_3)$, the tyre will work in the frictional area, which is unstable. In this case, if the brake pressure increases, the tire force and the utilized friction coefficient will decrease. The relationship between $\mu_x^t$ and $\lambda^t$ will change along the curve D to E, as shown in Figure 7. The estimated road peak friction coefficient is:

$$\mu_{x\max}^t = \mu_x^{t-1}$$

(44)

**The optimal slip ratio estimation**

Based on the extensive experimental road test, the rear-wheel anti-lock brake technique can be achieved when rear slip ratio is controlled during 0.1-0.3. The optimal slip ratio is the system's control target; it's related to road friction. According to the $\mu$-slip curve as shown in Figure 8, the relationship between the desired slip ratio $\lambda_d$ and the peak value of the road friction coefficient can be approximated by the following linear expression

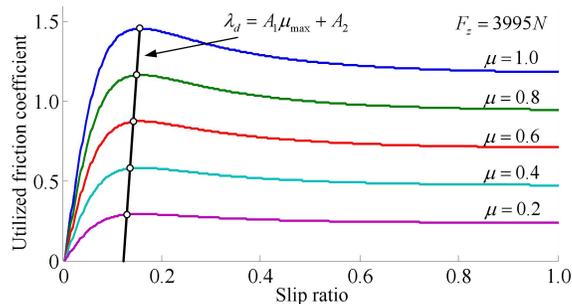

Figure. 8. Relationship between desired slip ratio and road friction

$$\lambda_d = A_1\mu_{\max} + A_2$$

(45)

where A1 and A2 is calibrated by experimental tire tests.

Using the least square method, values of A1 and A2 can be given as A1=0.05 and A2 =0.13, respectively.

To verify the effect of the proposed estimation method, assume that the actual tyre-road friction coefficient is known. The simulation results are shown in Figure (9). The actual road friction coefficient is switched from 0.2 to 0.8 at the time of 1.3s and from 0.8 to 0.5 at the time of 2.7s. It is found that the estimation friction coefficient is close to the actual value, so the estimation method is effective and the actual value can be estimated quickly and accurately. Based on these, the optimal slip ratio can be obtained timely in Figure (9).

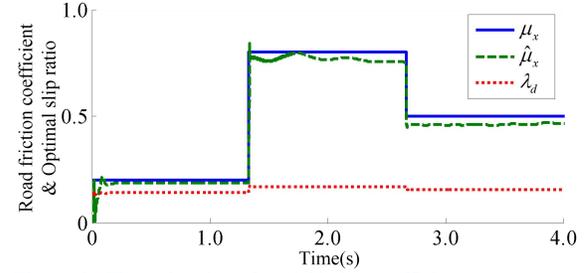

Figure. 9. The estimation of road friction coefficient

## 4. Sliding model controller designed

When the service brake system fails in the process of driving, the iEPB system is operated as a secondary brake. However, the slip ratio couldn't be controlled by iEPB, and it is difficult to be guaranteed the stability and security of vehicle. Therefore, sliding-mode control algorithm is proposed in this paper because of the presence of time-varying parameters and uncertainties of the inherent non-linearity system. [32] SMC has a strong adaptability based on the classical control theory of mathematics. [33, 34] Furthermore, it is more robust to variations in the parameters and to external disturbances comparing with other algorithms. [35, 36] Thus, it is suitable that SMC is applied in slip ratio control system.

The design process of SMC is introduced in detail in this section. And the complete anti-lock brake control system of rear wheel is described as shown in Figure 10.

In Figure 10, the vehicle model is constructed with Equations (1)-(25) based on the Magic Formula tyre model. The rear slip ratio is calculated using rear-wheel rotational speed and vehicle velocity which is estimated by the observer of ESC's controller. Then, the tyre/road friction coefficient estimator is proposed and the optimal slip ratio can be solved with $\mu-slip$ curve. The SMC as the upper controller is presented based on the tracking error between the actual rear slip ratio $\lambda_r$ and the optimal rear slip ratio $\lambda_{rd}$. Because of the estimated tyre/road friction coefficient, the variable road condition can be tracked timely and correctly. The control target of SMC is updated constantly. Then, the desired brake torque signal is output to make the tracking error slip ratio close to zero. In addition, another SMC as the lower controller is designed to control iEPB actuator. The desired brake torque is seen as the control target to make needed brake torque provided by lower iEPB actuator. And, the actual brake torque could be obtained by SMO. In this way, the maximum deceleration of vehicle can be reached and the best brake performance can be achieved by iEPB and control system.

## Upper controller

The design process of SMC as upper controller is introduced in detail as follows.

The tracking error $e_r$ could be calculated as

$$e_r = \lambda_r - \lambda_{rd} \tag{46}$$

The sliding surface $s_r$ is constructed using the tracking error.

$$s_r = e_r + c_1 \int_0^t e_r(\tau) d\tau \tag{47}$$

where $c_1$ is a system factor, it can control the tracking error and is adjusted based on the control performance

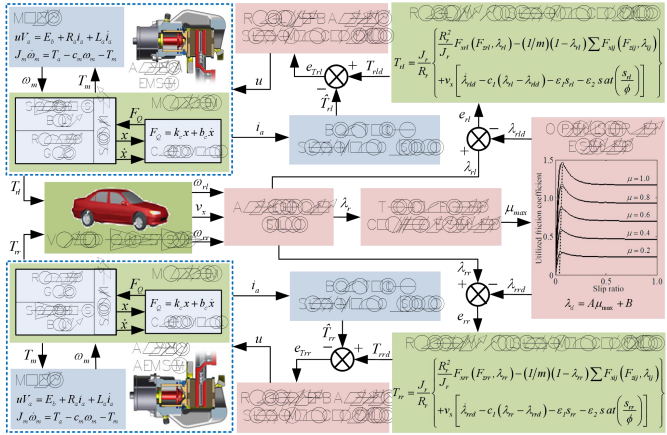

Fig. 10. Bock diagram of the system

The first derivative of $s_r$ is derived.

$$\dot{s}_r = \dot{e}_r + c_1 e_r = \dot{\lambda}_r - \dot{\lambda}_{rd} + c_1 e_r \tag{48}$$

The reaching law is

$$\dot{s}_r = -\varepsilon_1 s_r - \varepsilon_2 \operatorname{sgn}\left(\frac{s_r}{\phi}\right) \tag{49}$$

where $\varepsilon_1, \varepsilon_2$ are the sliding-mode gain factors, and they are determined by the stability of the sliding-mode controller and the Lyapunov stability theory.

The controller can induce undesirable high-frequency fluctuations. To ensure the performance of controller in practical application, the fluctuations should be suppressed by smoothing out the control discontinuity near the switching surface. Thus, the sign function in the controller is replaced by the saturation function.

The Equation (58) can be written.

$$\dot{s}_r = -\varepsilon_1 s_r - \varepsilon_2 sat\left(\frac{s_r}{\phi}\right) \tag{50}$$

The first differential of rear slip ratio is obtained from Equations (1), (3) and (7).

$$\dot{\lambda}_r = \frac{\omega_r R_r \dot{v}_x}{v_x^2} - \frac{\dot{\omega}_r R_r}{v_x} = \frac{(1-\lambda_r)\dot{v}_x}{v_x} - \frac{R_r}{v_x}\dot{\omega}_r$$

$$= \begin{bmatrix} -(R_r/J_r v_x)(R_r F_{xr}(F_{zr}, \lambda_r) - T_{epbr}) \\ +(1/mv_x)(1-\lambda_r)\sum F_{xij}(F_{zij}, \lambda_{ij}) \end{bmatrix} \tag{51}$$

With Equation (51), Equation (48) can be rewritten as

$$\dot{s}_r = \begin{Bmatrix} -(R_r/J_r v_x)(R_r F_{xr}(F_{zr}, \lambda_r) - T_{epbr}) \\ +(1/mv_x)(1-\lambda_r)\sum F_{xij}(F_{zij}, \lambda_{ij}) \\ -\left[\dot{\lambda}_{rd} - c_1(\lambda_r - \lambda_{rd})\right] \end{Bmatrix} \tag{52}$$

The control law of SMC for rear slip ratio is

$$T_{epbr} = \frac{J_r}{R_r}\begin{Bmatrix} \frac{R_r^2}{J_r}F_{xr}(F_{zr}, \lambda_r) - (1/m)(1-\lambda_r)\sum F_{xij}(F_{zij}, \lambda_{ij}) \\ +v_x\left[\dot{\lambda}_{rd} - c_1(\lambda_r - \lambda_{rd}) - \varepsilon_1 s_r - \varepsilon_2 sat\left(\frac{s_r}{\phi}\right)\right] \end{Bmatrix} \tag{53}$$

The stability of the designed SMC can be analyzed by the Lyapunov function as

$$V = \frac{1}{2}s_r^2 \tag{54}$$

The first derivative of $V$ is

$$\dot{V} = s_r \dot{s}_r$$

$$= s_r \begin{Bmatrix} -(R_r/J_r v_x)(R_r F_{xr}(F_{zr}, \lambda_r) - T_{epbr}) \\ +(1/mv_x)(1-\lambda_r)\sum F_{xij}(F_{zij}, \lambda_{ij}) \\ -[\dot{\lambda}_{rd} - c_1(\lambda_r - \lambda_{rd})] \end{Bmatrix}$$

$$= -s_r\left(\varepsilon_1 s_r + \varepsilon_2 \, sat\left(\frac{s_r}{\phi}\right)\right)$$

$$= -\varepsilon_1 s_r^2 - \varepsilon_2 |s_r|$$

(55)

If $\varepsilon_1 \geq 0, \varepsilon_2 \geq 0$, then

$$\dot{V} = -\varepsilon_1 s_r^2 - \varepsilon_2 |s_r| \leq 0$$

(56)

Therefore, the SMC system is stable.

**Lower controller**

The design process of SMC as lower strata controller is described in detail as follows.

When the piston moves forward or backward which means $\omega_m > 0$ or $\omega_m < 0$, the equation of brake torque is different. Based on that the armature inductance $La$ of the d.c. motor is very small and can be ignored, so the brake torque can be described with Equations (12)-(25).

$$\begin{cases} \omega_m \geq 0: \\ T_{epbr} = Q_3(i_{18}\eta_{89}\eta_n - \mu_s \sin\gamma)(Q_1 u - Q_2 \omega_m - J_n \dot{\omega}_m) \\ \omega_m < 0: \\ T_{epbr} = Q_3 \eta_n^{-1}(\mu_s \eta_n \sin\gamma - i_{18}\eta_{89})(Q_1 u - Q_2 \omega_m - J_n \dot{\omega}_m) \end{cases}$$

(57)

where

$$\begin{cases} \eta_n = \eta_{21}\eta_{53}\eta_{86}, Q_1 = \dfrac{k_t V_a}{R_a} \\ Q_2 = \dfrac{k_t k_e + c_n R_a}{R_a}, Q_3 = \dfrac{2\pi\mu_b r}{\pi d \mu_s \cos\gamma + P_h} \end{cases}$$

(58)

Because the brake torque is observed in system, the tracking error $e_T$ could be calculated as

$$e_T = \hat{T}_{epbr} - T_{epbrd}, \dot{e}_T = \dot{\hat{T}}_{epbr} - \dot{T}_{epbrd}$$

(59)

where $\hat{T}_{epbr}$ is the observed brake torque, the $T_{epbrd}$ is the desired brake torque, the control objective is to eliminate the tracking error between the actual brake torque and the desired one.

The sliding surface $s_T$ is constructed using the tracking error.

$$s_T = c_2 e_T + \int_0^t e_T(\tau) dt$$

(60)

where $c_2$ is a system factor, it can control the tracking error and is adjusted based on the control performance.

The first derivative of $s_T$ is derived.

$$\dot{s}_T = c_2 \dot{e}_T + e_T = c_2 \dot{e}_T + \hat{T}_{epbr} - T_{epbrd}$$

(61)

The reaching law is

$$\dot{s}_T = -\varepsilon_3 s_T - \varepsilon_4 sat\left(\frac{s_T}{\phi}\right)$$

(62)

where $\varepsilon_3, \varepsilon_4$ are the sliding-mode gain factors, and they are determined by the stability of the sliding-mode controller and the Lyapunov stability theory.

The Equation (70) can be rewritten as

$$\begin{cases} \omega_m \geq 0: \\ \dot{s}_T = \begin{bmatrix} c_2\left(\dot{\hat{T}}_{epbr} - \dot{T}_{epbrd}\right) - T_{epbrd} \\ +Q_3(i_{18}\eta_{89}\eta_n - \mu_s \sin\gamma)(Q_1 u - Q_2 \omega_m - J_n \dot{\omega}_m) \end{bmatrix} \\ \omega_m < 0: \\ \dot{s}_T = \begin{bmatrix} c_2\left(\dot{\hat{T}}_{epbr} - \dot{T}_{epbrd}\right) - T_{epbrd} \\ +Q_3 \eta_h^{-1}(\mu_s \eta_n \sin\gamma - i_{18}\eta_{89})(Q_1 u - Q_2 \omega_m - J_n \dot{\omega}_m) \end{bmatrix} \end{cases}$$

(63)

Thus, the control law of SMC for iEPB is

| Parameters | Value |
| --- | --- |
| Vehicle initial velocity ($v_o$) | 17 m/s |
| Vehicle mass (m) | 2100kg |
| Tire radius (R) | 327mm |
| Rotational inertia of the front/rear tire (J) | 1.7kg.m² |
| Height of the center of gravity (h) | 550mm |
| Distance between the front and rear axles (L) | 2800mm |
| Distance from the front axles to the center of gravity (a) | 1160mm |
| Distance from the rear axles to the center of gravity (b) | 1640mm |
| Brake disc friction coefficient (u) | 0.35 |
| Effective brake disc radius (r) | 200mm |

$$\begin{cases} \omega_m \geq 0: \\ u = \dfrac{\left\{\begin{array}{l}(Q_2\omega_m + J_n\dot{\omega}_m)\left[Q_3\left(i_{18}\eta_{89}\eta_n - \mu_s \sin\gamma\right)\right]^{-1} \\ +T_{epbrd} - c_2\left(\dot{\hat{T}}_{epbr} - \dot{T}_{epbrd}\right) - \varepsilon_3 s_T - \varepsilon_4 sat\left(\dfrac{s_T}{\phi}\right)\end{array}\right\}}{Q_1Q_3\left(i_{18}\eta_{89}\eta_1 - \mu_s \sin\gamma\right)} \\ \omega_m < 0: \\ u = \dfrac{\left\{\begin{array}{l}\eta_n(Q_2\omega_m + J_n\dot{\omega}_m)\left[Q_3\left(\mu_s\eta_n \sin\gamma - i_{18}\eta_{89}\right)\right]^{-1} \\ +T_{epbrd} - c_2\left(\dot{\hat{T}}_{epbr} - \dot{T}_{epbrd}\right) - \varepsilon_3 s_T - \varepsilon_4 sat\left(\dfrac{s_T}{\phi}\right)\end{array}\right\}}{Q_1Q_3\left(i_{18}\eta_{89}\eta_1 - \mu_s \sin\gamma\right)} \end{cases}$$

(64)

The stability of the designed SMC can be analyzed by the Lyapunov function as

$$\dot{V} = s_T \dot{s}_T$$
$$= -s_T\left(\varepsilon_3 s_T + \varepsilon_4 \, sat\left(\dfrac{s_T}{\phi}\right)\right)$$
$$= -\varepsilon_3 s_T^2 - \varepsilon_4 |s_T|$$

(65)

If $\varepsilon_3 \geq 0, \varepsilon_4 \geq 0$, then

$$\dot{V} = -\varepsilon_3 s_T^2 - \varepsilon_3 |s_T| \leq 0$$

(66)

Therefore, the SMC system is stable.

## 5. Simulation result and analysis

### 5.1 Description

To validate the performance of control strategy and the accuracy of parameters estimator, the brake process on different road conditions are simulated based on the built model in MATLAB/Simulink and AME-Sim. Assume that the vehicle is moving forward without steering input and the service brake, the brake torque is provided by only the iEPB system. Some related parameters are given in Table 3. In the following the simulation results, including parameters estimation and the control performance of the control algorithm, are discussed and analyzed in detail.

Table 3. The parameters of the tyre and the vehicle

### 5.2 Simulation results

**Comparison between SMC and PID**

To validate the performance of the SMC, a PID controller is designed in MATLAB/Simulink for comparison. Figure 11 shows the built model simulation results which the rear slip ratio is controlled by SMC and PID, respectively. The tyre/road friction coefficient is changed from 0.8 for the high friction coefficient to 0.2 for the low friction coefficient at 2s. Figure 11(1) and (2) show the vehicle velocity and rear-wheel velocity. And the desired rear slip ratio and the actual rear slip ratio are shown in Figure 11(3) and (4).

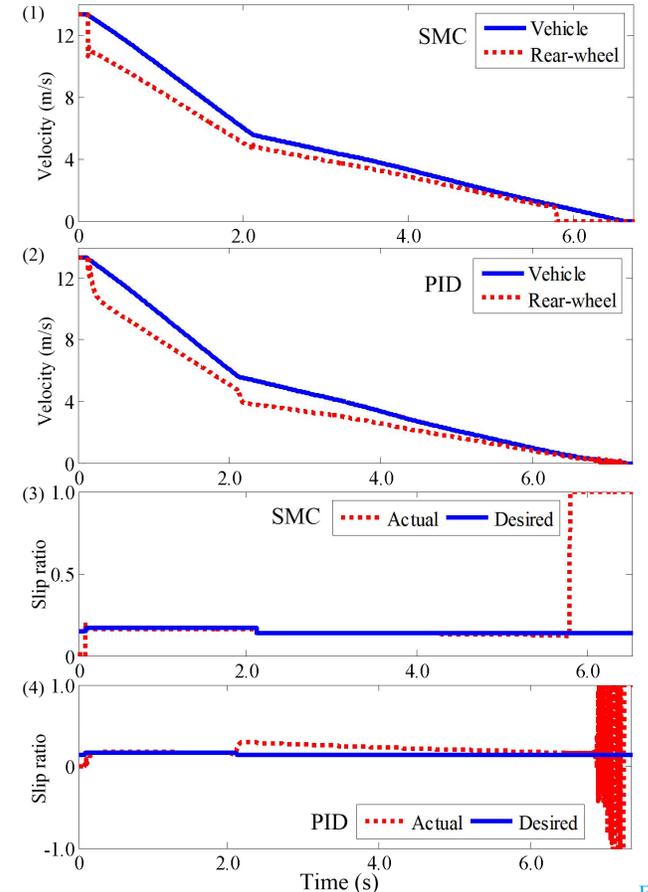

Figure. 11 The comparison of SMC's simulation results and PID's

In Figure 11(1), the vehicle velocity and the rear-wheel velocity reduce quickly during the first 2s, and then slow down when the road friction coefficient becomes low after 2s. The actual rear slip ratio tracks strictly the optimal value in Figure 11(3). The brake distance of vehicle is 31.44m by SMC. And the rear wheel is locked when vehicle velocity is less than 1m/s, it's normal and explicable. Because the ABS application is forced to quit when lower vehicle velocity in practice. However, in Figure 11(2) and (4), the wheel velocity reduces quickly in the road condition to change 0.8 to 0.2 at 2s. So the slip ratio error becomes large, it's unacceptable. Besides, the brake distance is 34.72m increased 3.28m using PID.

Based on the simulation results and above discussion, it is validated that the proposed SMC algorithm can control slip ratio precisely in different road condition and shorten brake distance comparing with PID. It's a well-known fact that PID algorithm is only suitable for certain dynamic parameters [37]. While the SMC algorithm can be widely used for non-linear systems with uncertain parameters and is robust to variations in the system parameters and to external disturbances. Thus, the SMC algorithm can be adopted in anti-lock brake control system.

*On single-friction-coefficient road*

In this part, the simulation on the single road condition is analyzed. The road conditions are as follows: the initial velocity is 50 km/h and the tyre/road friction coefficient is set as 0.8. The simulation results are shown in Figure 12.

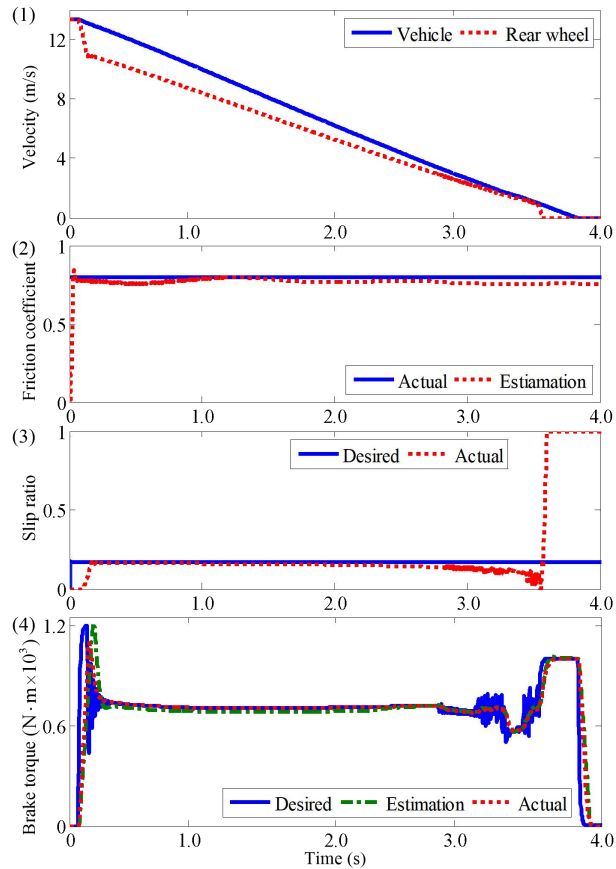

Figure. 12 The simulation results on single-friction-coefficient road

In Figure 12(1), the simulation results of vehicle velocity and the rear wheel velocity are shown. The actual tyre/road friction coefficient and the estimated friction coefficient by estimator are shown in Figure 12(2). The optimal slip ratio which is calculated timely using the estimated friction coefficient and the actual rear slip ratio are shown in Figure 12(3). In Figure 12(4), the desired brake torque which is the output signal of upper controller, the actual brake torque provided by iEPB actuator and the estimated brake torque by sliding-mode observer are shown.

In Figure 12(1), the vehicle velocity and the rear-wheel speed reduce gradually with time. The actual rear slip ratio is close to the desired slip ratio and the tracking error is less than 6.3% in the steady state in Figure 12(3). In Figure 12(2), the actual tyre/road friction coefficient curve and the estimated one are almost coincidence, and the error of them is less than about 5.2%. As for Figure 12(4), the actual brake torque tracks smoothly the desired value and the violent jitter of brake torque curve is eliminated. This characteristic of lower controller is fine when considering mechanical actuator. The iEPB system can adjust brake torque reposefully and the vibration of actuator can be weakened, even eliminated. The tracking error of brake torque is below about 7.8% in the steady state. There is delay when the brake torque is provided, because the clearance between friction plate and brake pad in iEPB actuator should be eliminated firstly. Besides, the estimation error of brake torque is less than 2.6% using SMO. As for the other simulation results, they are similar to that of SMC in Figure 11.

*On high-to-low-friction-coefficient road*

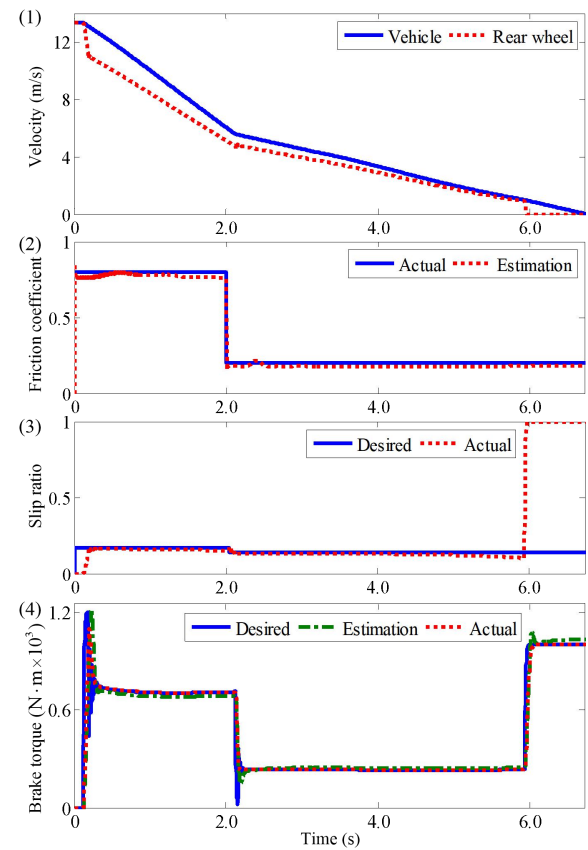

Figure. 13 The simulation results on high-to-low-friction-coefficient road

In order to study the applicability of the proposed vehicle anti-lock brake control algorithm, the control system simulation on high-to-low friction coefficient road is discussed in this part. The road condition is described with the tyre/road friction coefficient and the road condition is changed at 2s from 0.8 to 0.2. Other simulation parameters are the same as that previous simulation. The simulation results are shown in Figure 13.

As can be seen from Figure 13, the vehicle velocity and the rear-wheel velocity reduce quickly during the first 2s, and then slow down after 2s. When the tyre/road friction coefficient is changed suddenly, it is still estimated precisely by estimator. In addition, the brake torque decreases rapidly at 2s because of varying friction coefficient, and the actual brake torque by lower controller is smooth transition without vibration. The provided low brake torque is enough to adjust slip ratio. As for the other simulation results, they are similar to that in Figure 12.

***On a low-to-high-friction-coefficient road***

The control system is simulated on low-to-high friction coefficient road in this part. The road condition is described with the tyre/road friction coefficient, and the road condition is changed at 2s from 0.2 to 0.8. Other simulation parameters are the same as that previous simulation. The simulation results are shown in Figure 14.

In Figure 14(1), the simulation results of vehicle velocity and the rear wheel velocity are shown. The actual tyre/road friction coefficient and the estimated friction coefficient by estimator are shown in Figure 14(2). The optimal slip ratio and the actual rear slip ratio are shown in Figure 14(3). In Figure 14(4), the desired brake torque, the actual brake torque and the estimated brake torque by SMO are shown.

In Figure 14, the simulation results are similar as that on high-to-low friction coefficient road. The vehicle velocity and the rear-wheel velocity are reduced slowly during the first 2s, and then quick down when the road friction coefficient becomes high after 2s. The brake torque has to increase quickly in order to achieve anti-lock control when road condition is high friction coefficient.

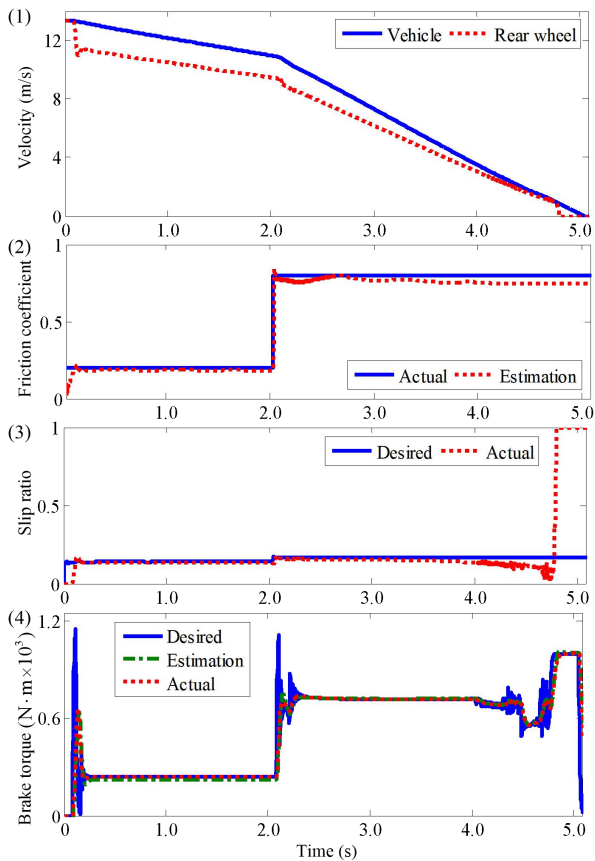

Figure. 14 The simulation results on low-to-high-friction-coefficient road

Thus, the designed anti-lock brake control algorithm is feasible, including upper and lower controller, the strong robustness and applicability of controller is verified based on above all simulation results. The load torque SMO and tyre/road friction coefficient estimator are effective.

# 6. Conclusions

For the slip ratio control when iEPB is operated as the secondary brake, this paper proposed rear-wheel anti-lock brake system for iEPB based on SMC algorithm. The vehicle and iEPB system were analyzed in detail and constructed in MATLAB/Simulink and AEM-Sim. The tyre/road friction coefficient estimator, SMO of load torque and SMC were investigated. Finally, the possibility of the proposed control system was verified by co-simulation on different road conditions. Conclusions could be drawn as follows.

a. The designed tyre/road friction coefficient estimator can make the estimation error below about 5.2%. SMO of the d.c. motor load torque has higher estimation accuracy than the existing observer and its estimation error is less than 2.6%. Thus, the actual parameters could be replaced by the values estimated and input the anti-lock brake system. The effective of estimator and observer is proved.

b. It is verified that the proposed SMC algorithm is feasible. In the steady state, it can make the rear slip ratio tracking error less than 6.3% and brake torque tracking error below 7.8%.

c. The iEPB is operated as the secondary brake in the process of driving when the conventional hydraulic brake fails. The rear slip ratio control could be achieved by the designed control system in this paper. The safety and stability of vehicle are improved significantly in specified condition. It provides theory basis for the development of anti-lock brake control system based on iEPB in practice.

# Funding

The authors greatly appreciate the support from National Key Science and Technology Projects (No.04) (Grant No: 2014ZX04002041) and the Fundamental Research Funds for the Central Universities.